# Precise measurement of energy of the first excited state of $^{115}$Sn ($E_{exc} \simeq 497.3$ keV)

V.A. Zheltonozhsky, A.M. Savrasov[1], N.V. Strilchuk, V.I. Tretyak

*Institute for Nuclear Research, 03028 Kyiv, Ukraine*

**Abstract**. Single beta decay of $^{115}$In to the first excited level of $^{115}$Sn ($E_{exc} \simeq 497.3$ keV) is known as β decay with the lowest $Q_\beta$ value. To determine the $Q_\beta$ precisely, one has to measure very accurately the $E_{exc}$ value. A sample of tin enriched in $^{115}$Sn to 50.7% was irradiated by proton beam at the U-120 accelerator of INR, Kyiv. The $^{115}$Sb radioactive isotope, created in $^{115}$Sn(p,n)$^{115}$Sb reaction, decays with $T_{1/2}$ = 32 min to $^{115}$Sn populating the 497 keV level with ≃ 96% probability. The total statistics of ~ $10^5$ counts collected in the 497 keV peak in series of measurements, exact description of the peak shape and precisely known calibration points around the 497 keV peak allowed to obtain the value $E_{exc}$ = 497.342(3) keV, which is the most precise to-date. This leads to the following $Q_\beta^*$ value for the decay $^{115}$In → $^{115}$Sn$^*$: $Q_\beta^*$ = 147 ± 10 eV.

## Introduction

While $^{115}$In isotope is present in natural mixture of elements with a big abundance of δ = 95.719(52)% [1], it is unstable in relation to beta decay. Nuclear decay $^{115}$In → $^{115}$Sn is characterized by one of the longest half-life for the observed single β$^-$ decays: $T_{1/2}$ = 4.41(25)×10$^{14}$ y [2, 3] (for the ground state to the ground state, g.s. to g.s., transitions). Until 2005, this process was considered as 100% g.s. to g.s., but in 2005 also decay to the first excited level of $^{115}$Sn was observed at the first time [4, 5] in measurements of a sample of metallic indium (mass of 929 g) during 2762 h with 4 HPGe detectors (≃ 225 cm$^3$ each) installed deep underground at the Laboratori Nazionali del Gran Sasso (Italy, 3600 m of water equivalent, m w.e.). The branching ratio of the decay was measured as b = (1.18 ± 0.31)×10$^{-6}$ that corresponds to partial half-life $T_{1/2}$ = (3.73 ± 0.98)×10$^{20}$ y [4]. The atomic mass difference $\Delta m_a$ between $^{115}$In and $^{115}$Sn (equal to energy release $Q_\beta$ in $^{115}$In β decay) at the time of measurements of Ref. [4] was known with quite low accuracy: $Q_\beta$ = 499 ± 4 keV [6]; however, in 2009 it was measured with extremely high accuracy of 10 eV: $Q_\beta$ = 497.489 ± 0.010 keV [7]. Taking into account that energy of the $^{115}$Sn first excited level was known as $E_{exc}$ = 497.334 ± 0.022 keV [3], the energy release in decay $^{115}$In → $^{115}$Sn$^*$ is equal $Q_\beta^*$ = 155 ± 24 eV that is the lowest known $Q_\beta$ value among observed β decays (the next one is $Q_\beta$ = 2.467 ± 0.002 keV for $^{187}$Re [8]). Decay scheme of $^{115}$In → $^{115}$Sn is shown in Fig. 1.

Observation [4] was confirmed in measurements of In sample (2566 g) at the HADES underground laboratory (Belgium, 500 m w.e.) with 3 HPGe detectors; slightly more precise values of half-life were obtained: $T_{1/2}$ = (4.1 ± 0.6)×10$^{20}$ y [9] and $T_{1/2}$ = (4.3 ± 0.5)×10$^{20}$ y [10].

As it was noted in [4, 5], β decay with such a low $Q_\beta$ value potentially can be used to limit (or to measure) the neutrino mass, in particular, looking for deviation of the energy β spectrum from theoretical shape; the last one was calculated in [11] taking into account that it is classified as 2-fold forbidden unique ($\Delta J^{\Delta\pi}$ = 3$^+$) and differs from the allowed shape. Half-life for the $^{115}$In → $^{115}$Sn$^*$ process was calculated in [9, 12] as a function of $Q_\beta^*$ value; however, for $Q_\beta^*$ = 155 ± 24 eV defined in [7] disagreement was by ~ 1 order of magnitude that probably is related with atomic effects which are poorly known yet at low energies and were not taken into account in the calculations.

---

[1] Corresponding author. Email address: asavrasov@kinr.kiev.ua

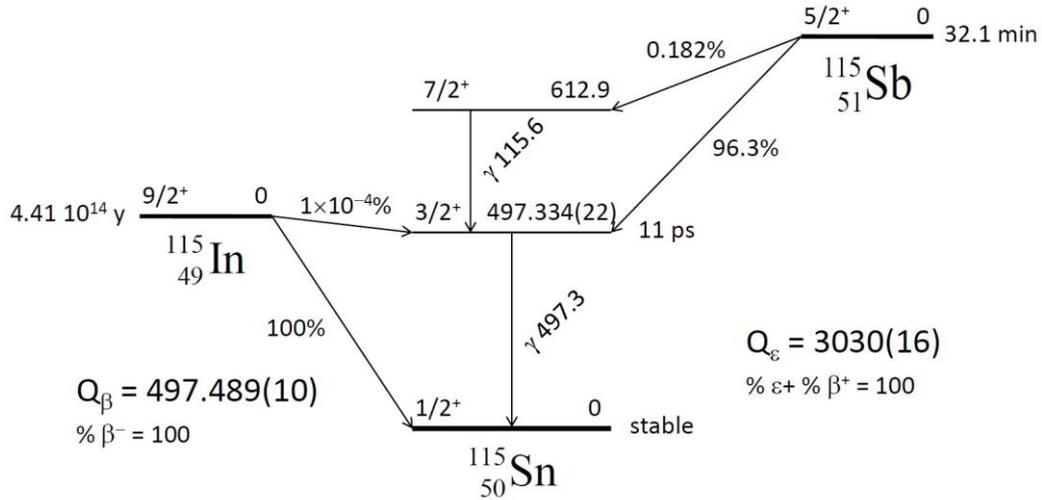

Fig. 1. Scheme of $^{115}$In decay and part of scheme of $^{115}$Sb decay with its main branch to the 497.3 keV level of $^{115}$Sn. Energies are in keV. Data are from [3, 7].

List of other potentially interesting candidates with low $Q_\beta^*$ values is given in [13, 14, 15, 16]; it should be noted that they are not observed yet, and $Q_\beta$ values are known with low accuracy of ~ 1 keV (or worse).

The uncertainty of 24 eV in $Q_\beta^*$ value is defined mainly by the 22 eV uncertainty in our knowledge of the energy of the $^{115}$Sn excited level: $E_{exc}$ = 497.334(22) keV [3]. The situation is even paradoxical to some extent: we know the absolute values of $^{115}$In and $^{115}$Sn masses of ~ 100 GeV (and their difference) with better accuracy than the level energy of ~ 0.5 MeV. It is clear that new, more exact measurement of energy of the first excited level of $^{115}$Sn is highly desirable.

Very recently this energy was precisely measured using $^{114}$Sn(n,γ)$^{115}$Sn reaction with cold neutrons and 30 mg Sn target enriched in 70% in the $^{114}$Sn isotope; the result is: $E_{exc}$ = 497.316 ± 0.007 keV, that leads to $Q_\beta^*$ = 173 ± 12 eV [17].

We present here results obtained in an alternative approach: by investigating decay of radioactive $^{115}$Sb ($T_{1/2}$ = 32 min) which was obtained in $^{115}$Sn(p,n)$^{115}$Sb reaction by irradiation of a sample of Sn. The $^{115}$Sb nuclide decays mainly ($\simeq$ 96%) to the 497 keV level of $^{115}$In that ensures an effective collection of statistics in the peak of interest.

**Experimental measurements**

The sample of tin enriched in $^{115}$Sn to 50.7% (the natural isotopic abundance of $^{115}$Sn is δ = 0.34% [1]) with mass of 0.23 g and thickness of 50 mg/cm$^2$ was irradiated by beam of protons with energy 6.8 MeV at the U-120 accelerator of the Institute for Nuclear Research, Kyiv (KINR). Intensity of the beam was equal 6.3×10$^{12}$ p·cm$^{-2}$·s$^{-1}$, and the cross-section for the $^{115}$Sn(p,n)$^{115}$Sb reaction is σ ~ 100 mb [18]. Energies of the emitted γ quanta were measured by HPGe detector (CANBERRA; 18% efficiency relatively to ∅3''×3'' NaI(Tl) detector at 1332 keV) which had energy resolution FWHM = 1.1 keV at E = 497 keV. The standard analogue CANBERRA and ORTEC electronics was used in the experiment. Counting rate varied from 20 to 100 counts/s during measurements.

For accurate determination of energy of the 497 keV γ's, we used calibration γ lines, not far from the energy of interest, those energies are known with high precision: $^7$Be ($T_{1/2}$ = 53 d) with $E_\gamma$ = 477.6035(20) keV; $^{124}$Sb ($T_{1/2}$ = 60 d) with $E_\gamma$ = 602.7260(23) keV; and $^{137}$Cs ($T_{1/2}$ = 30 y) with $E_\gamma$

= 661.657(3) keV [19, 20]. In addition, γ lines of $^{115}$Cd radioactive isotope ($T_{1/2}$ = 53 h) were used: $E_γ$ = 492.351(4) keV (which is very close to the 497 keV line of interest) and $E_γ$ = 527.901(7) keV [3]. $^{115}$Cd was produced in reaction $^{114}$Cd(n,γ)$^{115}$Cd with Cd target enriched in $^{114}$Cd to 99% (the natural abundance is 28.8%) at the KINR nuclear reactor. Strong annihilation line at 510.999 keV in the measured spectra was not used for calibration due to its bigger natural broadness but was used for additional control.

Measurements with the HPGe detector were performed for mixed $^7$Be + $^{115}$Cd + $^{124}$Sb + $^{137}$Cs + $^{115}$Sb source. After each 30 min, we added to the source additional $^{115}$Sb activity to keep stable accumulation rate of the 497 keV peak and dead time of the detector. Three irradiations of the Sn sample by p beam were performed, and 4 series of measurements with the HPGe detector were done after each irradiation. Totally 12 spectra were obtained, with statistics of (3–4)×10$^4$ counts in the 497 keV peak in each of them.

After each of the series, also longer measurements during 4 − 6 h were performed for estimation of non-linearity of the energy scale. Parts of the obtained spectra around the 497 keV region are shown in Fig. 2: measured during the first 30 min (left) and during 12 h (right).

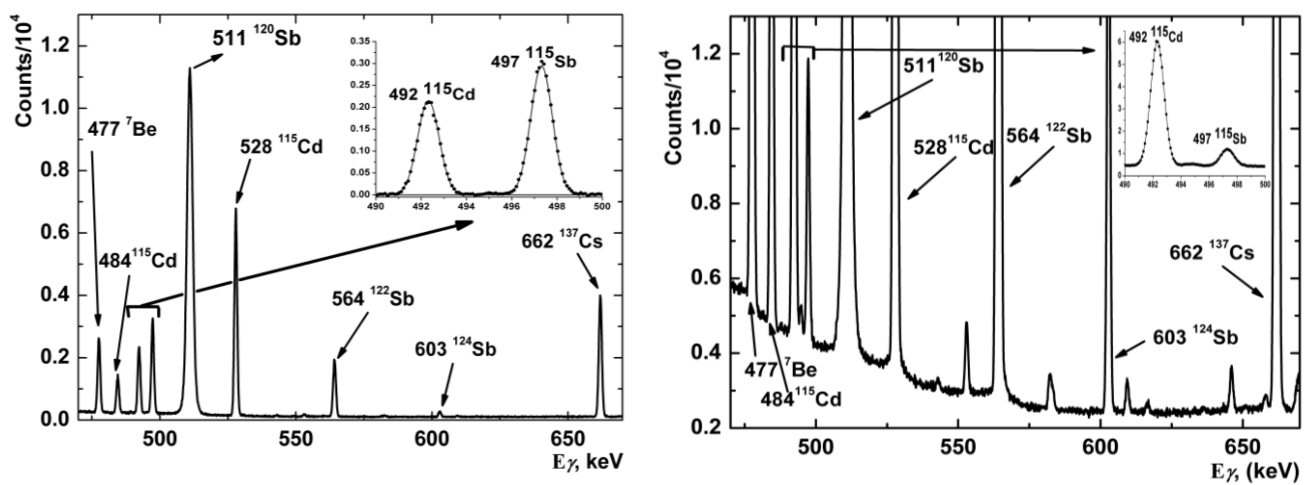

Fig. 2. Parts of energy spectra around the 497 keV region accumulated with the irradiated Sn sample during 30 min (left) and 12 h (right). Line at 484 keV results from the reaction $^{114}$Cd(n,γ)$^{115}$Cd. Lines with energies of 511 keV ($^{120}$Sb), 564 keV ($^{122}$Sb) and 603 keV ($^{124}$Sb) result from the (p,n) reaction on $^{120}$Sn, $^{122}$Sn and $^{124}$Sn, respectively; these isotopes were present in the used Sn target with abundance of: $^{120}$Sn (6.5%), $^{122}$Sn (1%) and $^{124}$Sn (1%).

**Results and discussion**

For accurate determination of energy of a peak, not only precisely known calibration points should be located near the peak of interest but also exact description of shapes of measured peaks and background under the peaks is extremely important. To fit the experimental spectra, we use the WinSpectrum code [21]. It describes the peak as the Gaussian with left and right tails which often appear due to incorrect regeneration of constant fraction of a spectrometer signal. Background is described by quadratic polynomial; a step present under each peak due to photoelectrons escape outside sensitive volume of detector is also taken into account. More details can be found in [21]. Fig. 3 shows part of one of the spectra with the fitted peaks as an example.

All the spectra were processed individually with their own individual backgrounds. Results for the obtained energies of the precisely known lines around the 497.3 keV peak of interest are given in Table 1.

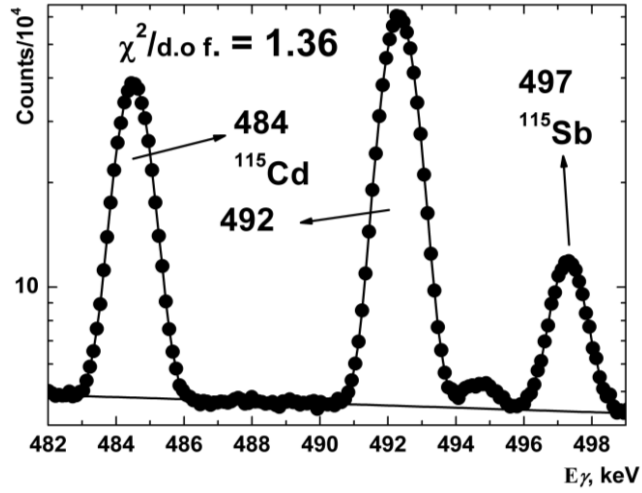

Fig. 3. Part of one of the spectrum with the fitted peaks in the region of the 497.3 keV peak of interest.

Table 1. Results for the obtained energies of lines around the 497.3 keV peak of interest for 12 individual spectra.

| | | Spectrum number | | |
|---|---|---|---|---|
| | | 1 | 2 | 3 |
| Nuclide | Table E (keV) | Fitted energy (keV) | | |
| $^{7}$Be | 477.6035(20) | 477.603(3) | 477.603(4) | 477.604(4) |
| $^{115}$Cd | 492.351(4) | 492.350(3) | 492.348(4) | 492.351(3) |
| $^{115}$Sb | 497.3 | 497.339(4) | 497.328(4) | 497.340(4) |
| $^{115}$Cd | 527.901(7) | 527.903(9) | 527.854(2) | 527.900(7) |
| $^{124}$Sb | 602.7260(23) | 602.723(3) | 602.715(16) | 602.717(7) |
| $^{137}$Cs | 661.657(3) | 661.656(3) | 661.656(4) | 661.655(4) |
| | | 4 | 5 | 6 |
| Nuclide | Table E (keV) | Fitted energy (keV) | | |
| $^{7}$Be | 477.6035(20) | 477.602(4) | 477.604(2) | 477.605(3) |
| $^{115}$Cd | 492.351(4) | 492.347(3) | 492.352(4) | 492.349(2) |
| $^{115}$Sb | 497.3 | 497.341(4) | 497.338(3) | 497.336(4) |
| $^{115}$Cd | 527.901(7) | 527.901(8) | 527.873(6) | 527.901(5) |
| $^{124}$Sb | 602.7260(23) | 602.718(13) | 602.722(14) | 602.716(9) |
| $^{137}$Cs | 661.657(3) | 661.658(2) | 661.656(2) | 661.658(2) |
| | | 7 | 8 | 9 |
| Nuclide | Table E (keV) | Fitted energy (keV) | | |
| $^{7}$Be | 477.6035(20) | 477.601(3) | 477.600(4) | 477.603(5) |
| $^{115}$Cd | 492.351(4) | 492.353(4) | 492.352(3) | 492.353(2) |
| $^{115}$Sb | 497.3 | 497.350(4) | 497.346(4) | 497.349(3) |
| $^{115}$Cd | 527.901(7) | 527.895(4) | 527.883(9) | 527.901(8) |
| $^{124}$Sb | 602.7260(23) | 602.725(6) | 602.714(15) | 602.719(9) |
| $^{137}$Cs | 661.657(3) | 661.659(4) | 661.655(3) | 661.654(4) |
| | | 10 | 11 | 12 |
| Nuclide | Table E (keV) | Fitted energy (keV) | | |
| $^{7}$Be | 477.6035(20) | 477.604(3) | 477.603(3) | 477.605(3) |
| $^{115}$Cd | 492.351(4) | 492.349(3) | 492.347(4) | 492.351(3) |
| $^{115}$Sb | 497.3 | 497.344(4) | 497.331(4) | 497.349(4) |
| $^{115}$Cd | 527.901(7) | 527.862(11) | 527.888(10) | 527.896(9) |
| $^{124}$Sb | 602.7260(23) | 602.722(7) | 602.721(8) | 602.722(5) |
| $^{137}$Cs | 661.657(3) | 661.658(3) | 661.655(4) | 661.660(2) |

In addition to the analytical description of the peaks, the experimentally measured shape of $^{7}$Be γ line at 477.6035 keV was also used in fitting procedures as a reference allowing to describe distortions of the peaks at 470 – 500 keV from the pure Gaussian in the same manner. Difference in positions of the γ lines, obtained by these two methods, was not greater than 1 eV.

To estimate a drift of the data acquisition system and non-linearity of the energy scale in the region of interest, we used two reference lines known with high accuracy: $E_γ$ = 477.6035(20) keV from $^{7}$Be and $E_γ$ = 661.657(3) keV from $^{137}$Cs (see Fig. 2). In measurements with the generated $^{115}$Sb, position of the $^{115}$Cd line $E_γ$ = 492.351(4) keV, located very close to our line of 497.3 keV, was controlled. In long (4 – 6 h) expositions without generated $^{115}$Sb, positions of lines $E_γ$ = 477.6035(20) keV, $E_γ$ = 661.657(3) keV and $E_γ$ = 492.351(4) keV were controlled. Geometry of targets in these expositions was the same as in measurements with $^{115}$Sb. In all the measurements, the shift of the energy scale was linear in the region of interest 477 – 662 keV, and non-linearity for $E_γ$ = 492.351(4) keV – and thus for our line too – was not greater than 1 eV.

Gamma line with energy $E_γ$ = 477.6035(20) keV was always used as the left reference point because of the lowest uncertainty of 2 eV and quite close location to our line of 497.3 keV. As the right reference point, different lines at $E_γ$ = 527.901(7) keV, $E_γ$ = 602.7260(23) keV and $E_γ$ = 661.657(3) keV were tested. However, line at 527.901(7) keV has relatively big table uncertainty of 7 eV, and $E_γ$ = 602.7260(23) keV – quite big error bars due to lower statistics in this peak. Thus for the final estimation of errors in the energy scale calibration, two lines were chosen: $E_γ$ = 477.6035(20) keV and $E_γ$ = 661.657(3) keV. Using these lines in 12 measurements, we obtained the weighted average uncertainty of 2 eV. For other pairs, this value was greater.

Difference in the determined energy of the 497.3 keV peak in different spectra was inside the error bars obtained in the fitting procedure, i.e. 3 – 4 eV (see Table 1). The weighted average energy and uncertainty after 12 measurements were calculated using the standard formulas:

$$<E> = \frac{\sum_{i=1}^{12} w_i E_i}{\sum_{i=1}^{12} w_i}, \qquad \frac{1}{\sigma_{<E>}^2} = \sum_{i=1}^{12} \frac{1}{\sigma_i^2}, \qquad w_i = \frac{1}{\sigma_i^2},$$

where $E_i$ and $σ_i$ are the energies and uncertainties in individual measurements given in Table 1. The obtained uncertainty was equal 1.1 eV. The total uncertainty of determination of the $^{115}$Sb line was calculated as square root of quadratically added: uncertainty of the line $E_γ$ = 477.6035 keV – 2 eV, the calibration uncertainty – 2 eV, and the weighted average statistical uncertainty – 1.1 eV, that results in the final value of 3 eV.

Similar procedures for processing the spectra were used in our previous work [22] for precise measurement (with 3 eV uncertainty) of energy of the first excited level of $^{197}$Au.

Fit of the measured experimental spectrum allowed to obtain the following value for the energy of $^{115}$Cd 492 keV peak (which is very close to our 497 keV peak of interest): $E_γ$ = 492.350(3) keV, which is in excellent agreement with the table value of 492.351(4) [3].

For the energy of the 497 keV γ line, the obtained result is: $E_γ$ = 497.341(3) keV. Correcting for the recoil of $^{115}$Sn nucleus, we calculate the energy of the first excited level of $^{115}$Sn: $E_{exc}$ = 497.342(3) keV. This value is in a good agreement with the evaluated data in Ref. [3]: 497.334(22) keV. However, it differs by 3.4σ from the recent result $E_{exc}$ = 497.316(7) keV [17].

Taking into account the $^{115}$In – $^{115}$Sn atomic mass difference of 497.489(10) keV [7], our value of $^{115}$Sn$^{*}$ $E_{exc}$ leads to the following $Q_β^{*}$ value for decay $^{115}$In → $^{115}$Sn$^{*}$: $Q_β^{*}$ = 147 ± 10 eV.

## Conclusion

Creating the $^{115}$Sb isotope in reaction $^{115}$Sn(p,n)$^{115}$Sb by irradiation of Sn sample by the proton beam and exactly measuring energy of the main γ quantum of 497 keV emitted in $^{115}$Sb decay, we obtained the following value for the energy of the first excited state of $^{115}$Sn: $E_{exc}$ = 497.342(3) keV. This is the most precise measurement to-date. It is in agreement with the evaluated data of 497.334(22) keV [3] but in 3.4σ disagreement with the recent result 497.316(7) keV [17].

The new value is obtained for the energy release in β decay of $^{115}$In to the first excited state of $^{115}$Sn: $Q_\beta^* = 147 \pm 10$ eV. This result leads to better agreement of the experimental partial half-life for this β decay with theoretical calculations of Ref. [12] but big difference still exists demanding to account new effects for $T_{1/2}$ calculations of β decays with extremely low $Q_\beta$ values.